\documentstyle[12pt,aaspp4]{article}
\newcommand \hii  {H~{\sc ii}}
\newcommand \hi  {H~{\sc i}}
\newcommand \ha   {H$\alpha$}
\newcommand \kms  {km~s$^{-1}$}
\newcommand \NH   {$N_{\rm H}$}

\begin{document}

\title{X-Rays from Superbubbles in the Large Magellanic Cloud.  V. The 
H\,II Complex N11}

\author{Mordecai-Mark Mac Low}
\affil{Max-Planck-Institut f\"ur Astronomie, K\"onigstuhl 17, 69117
Heidelberg, Germany \\E-mail: mordecai@mpia-hd.mpg.de}

\author{Thomas H. Chang, You-Hua Chu, Sean D. Points}
\affil{Dept. of Astronomy, Univ. of Illinois, 1002 W. Green St., Urbana, IL
61801, USA\\E-mail: thomas@juno.as.utexas.edu, chu@astro.uiuc.edu,
points@astro.uiuc.edu} 

\author{R. Chris Smith}
\affil{Dept. of Astronomy, Univ. of Michigan, 934 Dennison
Bldg., Ann Arbor, MI 48109, USA\\E-mail: chris@astro.lsa.umich.edu}

\and
\author{Bart P. Wakker}
\affil{Dept. of Astronomy, Univ. of Wisconsin,
475 N. Charter Street, Madison, WI 53706, USA\\E-mail: wakker@astro.wisc.edu}

%\slugcomment{version of 31 March 1997, M-MML}

\begin{abstract}
The large \hii\ complex N11 in the Large Magellanic Cloud contains OB
associations at several different stages in their life histories.  We
have obtained {\em ROSAT} PSPC and HRI X-ray observations, Curtis
Schmidt CCD images, echelle spectra in H$\alpha$ and [N II] lines, and
IUE interstellar absorption line observations of this region.  The
central bubble of N11 has an X-ray luminosity a factor of only 3--7
brighter than predicted for an energy-conserving superbubble, making
this the first detection of X-ray emission from a superbubble without
a strong X-ray excess.  The region N11~B contains an extremely young
OB association analogous to the central association of the Carina
nebula, apparently still embedded in its natal molecular cloud.  We
find that N11B emits diffuse X-ray emission, probably powered by
stellar winds.  Finally, we compare the tight cluster HD32228 in
N11 to R136 in 30~Dor.  The latter is a strong X-ray source, while the
former is not detected, showing that strong X-ray emission from
compact objects is not a universal property of such tight clusters.
\end{abstract}

\keywords{ISM: bubbles - Magellanic Clouds - X-ray: ISM}

\clearpage
\section{Introduction}

Giant \hii\ regions contain numerous massive stars.  In these
regions we expect that supernova remnants (SNRs) and stellar winds
from the most massive stars will interact with interstellar gas to
produce shocked, hot plasma that emits X-rays.  Indeed, bright 
diffuse X-ray emission has been detected in two nearby giant \hii\ 
regions: at levels of a few $\times 10^{35}$ erg s$^{-1}$ in the 
Carina Nebula (\markcite{sc82}Seward \& Chlebowski 1982; 
\markcite{cor94}Corcoran et al. 1994) 
and $\sim 10^{37}$ erg s$^{-1}$ in 
the 30 Doradus Nebula (\markcite{paperI}Chu \& Mac Low 1990, 
hereafter Paper I; \markcite{wh91a}Wang \& Helfand 1991a).  
Unfortunately, these giant \hii\ regions are so complex that it is 
difficult to understand their observed X-ray properties.

It has been suggested that stellar winds alone power the X-ray
emitting plasma in the Carina nebula, since no classical SNR
signatures exist, and since the diffuse X-ray morphology roughly
follows the distribution of early-type stars (\markcite{sc82}Seward \&
Chlebowski 1982; \markcite{wh82}Walborn \& Hesser 1982).  However,
interstellar absorption observations show the presence of
high-velocity gas at $\Delta$V = $\pm$200-300 km s$^{-1}$
(\markcite{wal82}Walborn 1982; \markcite{wh82}Walborn \& Hesser 1982),
and such large velocity offsets have otherwise only been seen in SNRs
(\markcite{jen76}Jenkins, Silk, \& Wallerstein 1976;
\markcite{fs83}Fitzpatrick \& Savage 1983).  Could there be hidden
SNRs producing both X-ray emission and high-velocity gas in the Carina
Nebula?

The diffuse X-ray emission in 30 Dor correlates well with its large
shell structures, reminiscent of the isolated X-ray bright
superbubbles described by \markcite{paperI}Paper I and by
\markcite{wh91b}Wang \& Helfand (1991b).  In these X-ray bright
superbubbles, the observed X-ray luminosities exceed by as much as an
order of magnitude the luminosities predicted by pressure-driven
bubble models (\markcite{wea77}Weaver et al.\ 1977;
\markcite{paperI}Paper I).  The source of the X-ray emission has been
proposed to be SNRs hitting the shell walls (\markcite{paperI} Paper
I; \markcite{wh91b}Wang \& Helfand 1991b) or mass-loaded SNRs within
the bubble (\markcite{ah96}Arthur \& Henney 1996).  We have also
identified superbubbles that are faint in X-rays, and found that the
upper limits on their X-ray luminosities are consistent with those
expected in pressure-driven bubble models (\markcite{paperIII}Chu et
al.\ 1995, hereafter Paper~III).  Can we convincingly detect X-ray
emission from superbubbles at levels predicted by pressure-driven
bubble models?

An {\em Einstein} High Resolution Imager (HRI) observation of the core
of 30 Dor showed two point-like sources, each having an X-ray
luminosity of nearly $10^{36}$ erg s$^{-1}$ (\markcite{wh91a}Wang \&
Helfand 1991a).  These two sources are projected near two tight
clusters of massive stars, R136 and R140 
(\markcite{ws89}Walborn \& Seward 1989); however, the luminosities
exceed those of stellar emission by at least 2--3 orders of magnitude.
Using recent {\em ROSAT} HRI observations, \markcite{wan95}Wang (1995)
identified Mk34 (10$''$ from R136) and R140 as their optical
counterparts and concluded that these X-ray sources are Wolf-Rayet +
black hole binaries.  Do such sources always occur in similar tight
clusters of massive stars?

To help answer these questions, we have studied the LMC \hii\
complex N11, the second-ranking \hii\ complex in the LMC
(\markcite{kh84}Kennicutt \& Hodge 1984).  N11 contains four 
OB associations: LH9, 10, 13, and 14 (\markcite{lh70}Lucke \& Hodge 
1970).  LH9 is surrounded by a large superbubble, while LH10, 13, 
and 14 are embedded in bright, compact \hii\ regions along the 
periphery of the superbubble, designated as the B, C, and E components 
of N11, respectively (see Figure~\ref{halpha}).  

We have selected N11 for three reasons.  First, the OB association
LH10 within N11B contains several O3 stars similar to those in the OB
associations in the Carina Nebula, but the surrounding \hii\ region
N11B has no high-velocity gas indicative of possible hidden SNRs.  If
the X-ray emission in the Carina Nebula comes only from hidden SNRs,
we would expect N11B to have a much lower X-ray luminosity.  Second,
the large superbubble surrounding LH9 in N11 forms a good candidate to
directly observe X-ray emission from a pressure-driven bubble.  Third,
the tight core of the association LH9, HD\,32228, contains at least
16 early-type stars with the principal components being WC5-6 and
O9.5II (\markcite{par92}Parker et al.\ 1992), analogous to R136 or
R140 in 30 Dor, and so can be examined for point-like X-ray sources.

N11 has recently been observed by \markcite{ros96}Rosado et al.\
(1996) using an imaging Fabry-Perot interferometer to map the velocity
structure in \ha\ and [O~{\sc iii}] lines.  Echelle spectroscopy in
\ha\ along a N-S line cutting through the main shell of N11 was
performed by \markcite{mea89}Meaburn et al.\ (1989).  \markcite{par92}
\markcite{par96} Parker et al.\ (1992, 1996) studied the stellar
content of the central associations of the main shell (LH9), and of
N11B (LH10), and \markcite{wp92}Walborn \& Parker (1992) compared N11B
to 30 Dor.  Maps of CO tracing molecular gas in the region have been
presented by \markcite{coh88}Cohen et al.\ (1988),
\markcite{idg91}Israel \& de Graauw (1991), and
\markcite{cal96}Caldwell (1996).

In the next section, we describe observations of N11 in the visible,
ultraviolet, and X-ray wavelengths that we have obtained in order to
better answer the questions raised above and understand the dynamics
of this region.  In \S3 we define the X-ray source regions and 
describe the analysis of the X-ray data.
In \S4 we derive the X-ray luminosity predicted by
the pressure-driven bubble model.  In \S5 we attempt to interpret the
observations of the various parts of the N11 complex, and finally in
\S6 summarize our conclusions.

\section{Observations} %2

We have observed N11 using a number of different techniques.
The distribution of the denser, photoionized gas in shell walls and 
molecular cloud surfaces can be seen in optical emission-line images 
that we obtained using a CCD camera on the Curtis Schmidt telescope 
at the Cerro Tololo Inter-American Observatory (CTIO).  In order
to trace the hot gas produced by shocks in stellar wind bubbles or
SNRs, we used pointed X-ray observations with the {\em ROSAT} 
Position Sensitive Proportional Counter (PSPC) and High Resolution 
Imager (HRI).  The kinematics of the photoionized gas are revealed 
by high-dispersion, long-slit spectra obtained with an echelle
spectrograph on the CTIO 4m telescope, and high-dispersion UV 
spectra obtained with the International Ultraviolet Explorer (IUE).
In this section, we describe these observations.

\subsection{Curtis Schmidt CCD images}

Multiple optical CCD images of two fields of the N11 complex were 
obtained with the Curtis Schmidt telescope at CTIO on 1995 January 28.
A front-illuminated Thomson 1024$\times$1024 CCD was used; its 19~$\mu$m 
pixel size corresponded to 1\farcs835 pixel$^{-1}$.  The field-of-view 
was 31\farcm3.  Narrow-band filters of H$\alpha$ ($\lambda_c= 6564$~\AA,
$\Delta\lambda = 20$~\AA), [S~{\sc ii}] ($\lambda_c = 6724$~\AA, 
$\Delta\lambda = 50$~\AA), and [O~{\sc iii}] ($\lambda_c = 5010$~\AA,
$\Delta\lambda = 50$~\AA) were used to isolate diagnostic emission lines.
Broader red and green filters ($\lambda_c = 6840$~\AA, $\Delta\lambda =
95$~\AA\ and $\lambda_c = 5133$~\AA, $\Delta\lambda = 92$~\AA, 
respectively) were used to obtain emission-line-free continuum images,
which provide templates for continuum subtraction from the emission-line
images.  The data were reduced using the IRAF\footnote{IRAF is distributed
by the National Optical Astronomy Observatories (NOAO).} software package 
for bias subtraction, flat-fielding, and sky subtraction.  The images of the 
two fields were mosaicked after multiple frames in each field were shifted 
and combined.  The total exposure times for each position in the final 
mosaicked images are 2700 s in H$\alpha$, 4800 s in [S~{\sc ii}], 3600 s in 
[O~{\sc iii}], 1800 s in the red continuum, and 1500 s in the green 
continuum.  Figure~\ref{halpha}(a) presents the green continuum image to 
show the location of stars, and Figure~\ref{halpha}(b) the 
continuum-subtracted H$\alpha$ image to show the distribution of 
ionized gas.

The flux calibration of the CCD images was made using observations of 
photometric standard stars.  The sky-subtracted \ha\ fluxes of N11 
determined from the CCD images agree within 10\% uncertainty with those 
determined from the photoelectrically calibrated PDS scans of 
\markcite{kh86}Kennicutt \& Hodge's (1986) photographic plates, kindly 
provided to us by Dr.\ R. C. Kennicutt.  The \ha\ fluxes are converted 
to emission measures, assuming an electron temperature of 10$^4$ K.  
The faintest nebulosities detected by the deep CCD images have emission 
measures of $\sim10$ cm$^{-6}$~pc; the brightest parts have emission
measures over 10$^5$ cm$^{-6}$~pc.

\subsection{ROSAT X-ray Observations}

We used the {\em ROSAT} PSPC and HRI to obtain X-ray observations of
N11.  The PSPC is 
sensitive to the energy range of 0.1--2.4 keV, while the HRI is
sensitive to 0.1--2.0 keV.  The PSPC has an energy resolution of
$\sim$45\% and an on-axis angular resolution of $\sim$30$''$ at 1 keV.
The HRI has a very poor energy resolution but a higher angular
resolution, $\sim$5$''$.  Detailed information about the performance
and characteristics of these instruments can be found in Pfeffermann
\markcite{pfe86} et al.\ (1986) or the \markcite{rosat}{\em ROSAT}
Mission Description (1991).

The PSPC observation of N11 was numbered rp900320.  A total
integration time of 29.65 ks was obtained, with 16.7 ks between 1992
November 16 and 1992 December 4 and 12.9 ks during 1993 February
14--15.  The PSPC was operated in the low-gain state throughout the
observation.  The HRI observation of N11 was numbered rh900321.  A
total integration time of 27.1 ks was obtained between 1992 October 25
and 1993 February 18.  

The PSPC observation is best suited for revealing diffuse X-ray
emission.  To maximize the S/N ratio without significantly degrading
the spatial resolution, we have binned the PSPC image, which is
heavily oversampled compared to its instrumental resolution, by a
factor of 10 to obtain a pixel size of 5$''$ pixel$^{-1}$, and
smoothed the image with a gaussian of $\sigma$ = 5 pixels, or 25$''$.
The smoothed PSPC image in the 0.1--2.4 keV band is shown in
Figure~\ref{rawx}.  To show the relationship between H$\alpha$ and
X-ray emission, Figure~\ref{hax} presents an \ha\ image overlaid by
X-ray contours derived by binning the PSPC image by a factor of 8 and
smoothing it with a gaussian of $\sigma$ = 2 pixels, or $8''$.

The HRI observation is better suited for revealing point sources. 
For a 6$''$-diameter detection cell, no point sources are detected 
in N11 and its vicinity at $\ge 3\sigma$ levels, or 2.3$\times$10$^{-4}$ 
counts s$^{-1}$.  This count rate corresponds to a luminosity of
4$\times10^{33}$ erg s$^{-1}$, using the conversion factor for 
a 5$\times$10$^6$ K plasma and an absorption column density of
log $N_{\rm H}$ = 21.5 in Figure 10.3 of the $\em ROSAT$ Mission
Description (1991).

The SNR N11L (the bright source on the right) is the only obvious
X-ray source in the 
HRI image.  We have binned the HRI image by a factor of 10, resulting in 
a pixel size of 5$''$ pixel$^{-1}$, and smoothed it with gaussians of 
$\sigma$ = 50$''$ -- 100$''$.  Regions with bright diffuse X-ray emission 
detected by the PSPC are also detected in the heavily smoothed HRI images.
Since these heavily smoothed HRI images have effective resolutions 
comparable to or lower than the PSPC's resolution, and since the smoothed 
HRI image has lower S/N ratios than the PSPC image, the HRI image is not 
shown in this paper.  Analysis of the HRI image of N11L will be reported 
in another paper.

\subsection{Long-Slit Echelle Spectroscopy of Emission Lines}

\label{ech}
We obtained high-dispersion spectra of N11 using the echelle
spectrograph with the long-focus red camera on the 4m telescope at
CTIO.  A 79 l/mm echelle grating was used.  We used the instrument 
as a single-order, long-slit spectrograph by inserting a postslit 
interference filter and replacing the cross-dispersing grating with 
a flat mirror, so that a single echelle order around H$\alpha$ and 
[N~{\sc ii}] $\lambda\lambda$6548, 6583 could be imaged over the entire 
slit.  The data were recorded with a Tek 2048$\times$2048 CCD using Arcon 
3.6 controller.  The pixel size was 24 $\mu$m pixel$^{-1}$, which provided 
a sampling of 0.082~\AA\ (3.75 \kms) along the dispersion and 
0\farcs267 on the sky.  Spectral coverage was effectively limited by 
the bandwidth of the interference filter to 125~\AA.  Coverage along 
the slit was 245$''$, including an unvignetted field of $\sim$ 3\farcm5.
The instrumental profile, as measured from Th-Ar calibration lamp lines, 
was 16.1 $\pm$ 0.8 \kms\ FWHM at the \ha\ line.  This range
includes variations in the mean focus and variations in the focus over 
a single spectrum.  Spatial resolution was determined by the seeing, 
which was $\sim 1''$.  The exposure times were 10--20 minutes.

The data were bias subtracted at the telescope, and later processed 
using the IRAF software for flat-fielding, wavelength calibration, 
distortion correction, and removal of cosmic rays.  Wavelength 
calibration and distortion correction were performed using Th-Ar lamp 
exposures taken in the beginning of each night.  To eliminate velocity 
errors due to flexures in the spectrograph, the absolute wavelength 
scale in each spectrum was referenced to the geocoronal H$\alpha$ 
line at zero observed velocity.

Three slit positions were observed in N11.  One E-W oriented slit
was centered on the star Sk$-66^\circ31$ (RA
04$^{\rm h}$56$^{\rm m}$40$^{\rm s}$, Dec $-66^\circ33'33''$, J2000) 
near the center of the main shell in N11.  The other two slit positions 
were centered on the star PGMW~3021 (\markcite{par92}Parker et al.\ 
1992) in N11B $\sim244''$N of Sk$-66^\circ31$, with one slit E-W 
oriented and the other N-S oriented. 
The slit positions are marked in Figure~\ref{halpha}(b) and the
echellograms are presented in Figure~\ref{echimage}.
The spectral features detected 
in the echellograms (Figure~\ref{echimage}) include the \ha\ line and 
the [N~{\sc ii}] $\lambda\lambda$6548, 6583 lines, in addition to the 
faint telluric airglow features of OH.  The [N~{\sc ii}] lines are 
too weak to be of much use.

\subsection{High-dispersion IUE Observations}

High-dispersion IUE spectra are available for two stars in N11:
SWP\,15065 for HD~32228 (Sk$-68^\circ$28) and SWP\,49304 for 
Sk$-66^\circ$31.  The spectrum of HD~32228, retrieved from the IUE 
archives and presented by \markcite{chu94}Chu et al.\ (1994), shows 
C~{\sc iv} and possibly Si~{\sc iv} interstellar absorption at 
$\sim270$ \kms, and S~{\sc ii} and Si~{\sc ii} at $\sim290$ \kms.  
The spectrum of Sk$-66^\circ$31, an O9I star, was obtained by us in 
the 16th Episode of IUE.  Unfortunately, this spectrum is too noisy 
to be very useful, although by much smoothing one can find features 
that could be Magellanic C~{\sc iv} and possibly Si~{\sc iv}
absorption.

\section{Analysis of X-ray Data}

To analyze the PSPC X-ray data, we first define source regions
(detailed in \S 3.1), then extract the count rate and spectral energy
distribution for each region.  The observed spectral energy
distribution is a convolution of the intrinsic spectrum, foreground
absorption, and the instrumental response.  The instrumental response
of the ROSAT PSPC is well calibrated.  It is possible to fold the
instrumental response into plasma emission models (e.g.,
\markcite{rs77}Raymond \& Smith 1977) and fit the results to the
observed spectral energy distribution in order to determine the plasma
temperature $kT$, foreground absorption $N_{\rm H}$, and X-ray
luminosity $L_x$.  We can express the foreground absorption in terms
of hydrogen column density $N_{\rm H}$ by using the effective
cross-sections per hydrogen atom of \markcite{mm83}Morrison \&
McCammon (1983).  The goodness of our spectral fits is shown by
$\chi^2$ grid plots in the $kT - N_{\rm H}$ plane.  Using a 30\%
cosmic abundance that is appropriate for the LMC, we have made
spectral fits to the observed spectral energy distribution
(Figure~\ref{x-spec}) and generated $\chi^2$ grid plots
(Figure~\ref{gridplots}).  The best-fit $kT$, log $N_{\rm H}$, and
$L_x$ are listed in the top section of Table 1.

The number of counts received from our sources is usually too low to
constrain the spectral fits well.  Therefore, we use other independent
estimates of foreground absorption to confine the plasma temperature
fits.  The various methods we use to estimate absorption columns are
described in detail below in \S~3.2.  The plasma temperature $kT$ and
the absorption column $N_{\rm H}$ are then used to determine the
intrinsic X-ray fluxes and luminosities.  The best independently
estimated log $N_{\rm H}$, the corresponding best-fit $kT$, and the
resultant $L_x$ are listed in the bottom section of Table 1.

When absorbing column densities are of order $10^{22}$ cm$^{-2}$ or
higher, emission below about 0.5~keV is almost completely absorbed,
even though {\em ROSAT} is sensitive down to 0.1 keV.  The intrinsic
luminosity in the 0.1--2.4 keV band is then dominated by the
absorption correction in the 0.1--0.5 keV range, but virtually no
photons are detected in this range, giving very uncertain results.  We
do give the luminosities derived for both 0.1--2.4 keV and 0.5--2.4
keV bands in Table 1.  Although we used the 0.1--2.4 keV band to find
conservative upper limits for X-ray dim superbubbles in Paper III, the
high luminosities given in Table 1 from considering this full band in
regions with high absorption appear unphysical.  The luminosities
derived from the 0.5--2.4 keV band are more physically meaningful.

\subsection{Selection of Source Regions}

There is excellent correspondence between the X-ray emission and
interstellar structures revealed in the \ha\ image.  Frequently,
enhanced X-ray emission is seen within large shells.
Guided by the \ha\ image, we divide the X-ray emission from N11 into 
seven source regions, as shown in Figure~\ref{hax} and listed in 
Table~\ref{spectra}.  Four of these regions are associated with 
large shells.  We define ``Shell 1'' as the main shell of N11, 
encompassing LH 9.  ``Shell 2'' is the faint shell structure $10\arcmin$ 
north of the main shell.  ``Shell 3'' is the small shell structure NW 
of Shell 1 and SW of Shell 2.  ``Shell 4'' is the shell located $10'$ 
south of the main shell.  Two source regions are selected to 
include the B and C components of N11, respectively.  We notice faint
H$\alpha$ filaments correlated with diffuse X-ray emission extending
NW of Shell 2 suggesting a blowout into a lower density region which
we define as the ``NW diffuse'' source region.  The X-ray data of
these regions are analyzed individually.

\subsection{Independent Estimates of Absorption Columns}

Soft X-rays are absorbed by all phases of the interstellar medium (ISM).
We have used two methods to estimate the ranges of
absorption column densities.  As described below, both methods
suffer from large uncertainties.

In the first method we limit the possible range in absorption column
$N_{\rm H}$ by using previous observations of atomic and molecular
gases and our current observations of ionized gas toward N11.  An \hi\
survey of the LMC by \markcite{roh84}Rohlfs et al. (1988) has shown a
range of LMC \hi\ column density of (1.8--2.9) $\times 10^{21}$
cm$^{-2}$ in the four pointings near N11, with a beam size of 15$'$.
A Galactic foreground H~{\sc i} column density of \NH\ = $(5.6 \pm
2.0) \times 10^{20}$ cm$^{-2}$ was derived by \markcite{si91}Schwering
\& Israel (1991) using the large-scale \hi\ survey by
\markcite{cle79}Cleary, Heiles \& Haslam (1979). 

A high-resolution ESO-SEST survey shows a molecular cloud complex
roughly encircling the N11 complex.  Except for N11B, X-ray emission
is not observed from any regions containing CO, possibly due to
obscuration by the molecular clouds.  The clouds show a peak CO intensity,
I$_{\rm CO}$ = 20 K \kms (\markcite{is93}Israel et al.\ 1993).  To
translate this into a column density of H$_2$, we scale the galactic
CO/H$_2$ conversion ratio $X_{\rm MW}=N_{\rm H_{2}} / W_{\rm CO}=2.3
\times 10^{20}$ cm$^{-2}$~K$^{-1}$~km$^{-1}$~s
(\markcite{blo89}Bloemen 1989) by the LMC metallicity of 1/3.  The
observed CO intensity then correspond to a $N_{{\rm H}_2}$ of $\sim 1
\times 10^{21}$ cm$^{-2}$.

Using our calibrated CCD \ha\ images, we find emission measures of $4
\times 10^4$ cm$^{-6}$ pc in N11B and less than $1 \times 10^3$
cm$^{-6}$ pc in shells and other low-surface brightness areas.
Assuming the depth of the \ha\ emission region is as large as the
width of the emission region, we estimate ionized hydrogen column
densities of $\sim 3.5 \times 10^{21}$ cm$^{-2}$ in N11B and $\sim 2
\times 10^{20}$ cm$^{-2}$ in shells and low-surface brightness areas.

By combining these values we can estimate upper and lower limits to
the absorbing hydrogen column density.  For the lower limit, we assume
that N11 is in front of all the LMC~\hi\, and that no more than half
of the ionized gas in N11 contributes to the absorption column, so
that the lower limit is the sum Galactic~$N_{\rm HI} + \frac12 \mbox{
LMC } N_{\rm HII}$.  We use the appropriate \hii\ column densities for
the different regions as given above.  For N11B this yields a lower
limit of $\log N_{\rm H} = 21.3$, and $\log N_{\rm H} = 20.9$ for the
other regions.
For the upper limit, we examine the CO image of
\markcite{cal96} Caldwell (1996).  
In N11B, the only X-ray emitting region that
coincides with strong CO emission, we use 
the sum Galactic~$N_{\rm HI} + \frac12 (\mbox{ LMC } N_{\rm HII} +
\mbox{ LMC } N_{\rm HI} +2 \mbox{ LMC } N_{{\rm H}_2})$ as an upper
estimate, while in regions lacking CO emission we 
neglect the H$_2$ contribution entirely.  This gives an upper limit of
$\log N_{\rm H} = 21.6$ for N11B, and $\log N_{\rm H} = 21.3$ for the
other regions. 

The second method to estimate absorption column uses optical
extinction of stars embedded in the X-ray source regions.  An overall
average reddening of $E(B-V)= 0.05$ mag has been found in the stars of
LH~9, the OB association in the main N11 shell, and $E(B-V)=0.17$ mag
in LH 10 in N11 B (\markcite{par92}Parker et al.\
1992). \markcite{bru75}Brunet (1975) has reported an average Galactic
foreground reddening of $E(B-V)= 0.07$ mag. However, small scale
variations in Galactic reddening ranging from 0.0 to 0.15 mag toward
the LMC have been found by \markcite{oes95}Oestreicher et al.\ (1995)
using a more complete sample of 1500 Galactic foreground stars.  We
adopt a Galactic gas-to-dust ratio of $N_{\rm H}/E(B-V)=4.8 \times
10^{21}$ cm$^{-2}$~mag$^{-1}$ (\markcite{boh78}Bohlin et al.\ 1978),
and an LMC gas-to-dust ratio of $2.4 \times 10^{22}$
cm$^{-2}$~mag$^{-1}$ (\markcite{fit86}Fitzpatrick 1986).  Adopting the
reddening values of \markcite{par92}Parker et al.\ (1992) we estimate
$20.3 \leq \log N_{\rm H} \leq 21.1$ for the main shell and $21.1 \leq
\log N_{\rm H} \leq 21.6$ for N11B.  By the same arguments, we can
make an independent estimate of $20.9 \leq \log
N_{\rm H} \leq 21.6$ using the LMC intrinsic reddening map of
Oestreicher \& Schmidt-Kaler (1996).  

These values agree well with our
estimates from the emission-line observations of the ISM described
above. 
We use a combination of the two estimates to constrain the
absorption column densities, taking the lower of the two upper limits
and the upper of the two lower limits to give the most likely range of
$N_{\rm H}$.   The range of $N_{\rm H}$ that we have actually
used for each region is listed in the bottom section of Table 1.

\section{X-ray Emission from Pressure-Driven Bubble Model} %3

To predict the X-ray luminosity from a pressure-driven bubble we
follow the procedure described in \markcite{paperI} Paper~I, as
corrected in \markcite{paperIII} Paper~III.  By assuming that
conductive evaporation determines the interior structure of the bubble
\markcite{wea77} (Weaver et al.\ 1977), we can compute the X-ray
luminosity if we know the radius of the bubble in pc, $R_{pc}$, its
expansion velocity in km s$^{-1}$, $v_5$, and the emission measure in
H$\alpha$ on a tangential path through the shell in cm$^{-6}$ pc,
$EM$.  If we assume that the thickness of the shell, $\Delta R \ll R$,
the electron temperature $T_e \simeq 10^4$, and the mean atomic mass
of the ambient medium $\mu_a = (14/11) m_H$, 
we can use equations~(7--10) from Paper~III to derive the X-ray
luminosity directly from the three observables,
\begin{equation}
L_x \simeq (6.7 \times 10^{29}\mbox{ erg s}^{-1}) \xi I EM^{5/7}
R_{pc}^{12/7} v_5^{1/7},
\label{Lx}
\end{equation}
where $\xi$ is the metallicity, which we take to be 0.3 in the LMC, and
$I$ is a slowly varying function of the interior temperature of the bubble
which we can take to have value $I \sim 2$.  The external density into
which the bubble is expanding is then
\begin{equation} \label{n0}
n_0 \simeq (22 \mbox{ cm}^{-3}) EM^{1/2} v_5^{-3/2} R_{pc}^{-1/2}, 
\end{equation}
the age of the bubble in Myr is
\begin{equation} \label{t6}
t_6 = 0.59 R_{pc} / v_5, 
\end{equation}
and the mechanical luminosity $L_{\rm mech}$ from stellar winds and
SNRs driving the bubble is
\begin{equation} \label{mech}
L_{mech} \simeq (8 \times 10^{30} \mbox{ erg s}^{-1}) EM^{1/2} v_5^{3/2}
R_{pc}^{3/2}. 
\end{equation}
The kinetic energy $K$ of the shell in a pressure driven bubble can be
derived by noting that the thermal energy of the interior is $U = (5/11)
L_{\rm mech} t$ \markcite{wea77} (Weaver et al.\ 1977), and that $K +
U = 50/77 L_{\rm mech} t$ (Mac Low \& McCray 1988) \markcite{mm88}, so
\begin{equation} \label{ke}
K = (15/77) L_{\rm mech} t = (2.9 \times 10^{43} {\rm ergs}) EM^{1/2}
v_5^{1/2} R_{pc}^{5/2}.
\end{equation}
(Alternatively, the final result can be derived by taking $K = (1/2) M
v_s^2$, where the mass of the shell $M = (4/3) \pi R^3 \mu_a n_0$.)

\section{Interpretation} %4

\subsection{Shell 1}

In \markcite{paperI} Paper~I we showed that some LMC superbubbles are
X-ray bright, by which we mean that they have X-ray luminosities much
higher than predicted by the pressure-driven bubble model, while in
\markcite{paperIII} Paper~III we placed upper limits on the X-ray
luminosities from several other superbubbles that were close to the
prediction of the model. Here we derive the X-ray luminosity of
Shell~1 and show that it is only slightly X-ray bright, with a
luminosity a factor of 3--7 brighter than predicted by
pressure-driven bubble model.

X-ray emission from the central region of N11 lies within the optical
shell seen in H$\alpha$.  Performing spectral fits using free
parameters in this region shows a 99\% confidence range in absorption
and plasma temperature of $\log N_{\rm H} = $20.5--22.0 and
$kT=$0.1--0.4~keV (see Table~\ref{spectra}).  This absorption column
density is consistent with the values we derived in \S~3.2.  If we
constrain the column density to those values, we can derive an X-ray
luminosity of (3.2--4.0)$ \times 10^{35}$ erg s$^{-1}$ in the 0.5--2.4
keV band, assuming a distance of 50 kpc to the LMC \markcite{pan91}
(Panagia et al. 1991).

The velocity structure of the main shell is quite complex.
Figure~\ref{echimage}(a) shows an echellogram from an E-W slit across the
central region, while Figure~\ref{echmain} shows velocity profiles at four
different points on that slit.  The broadest splitting seen is some 120
km~s$^{-1}$ (Fig.~\ref{echmain}a), suggesting an expansion
velocity as high as 60~km~s$^{-1}$.  \markcite{mea89} Meaburn et al.\
(1989) observed the H$\alpha$ line with an echelle spectrograph along
a N-S line passing through HD~32228, crossing our slit just west of
center.  They find as many as five velocity components in the central
region, which they attribute to multiple expanding shells in a
cellular structure.  \markcite{ros96} Rosado et al.\ (1996) performed
Fabry-Perot observations of the entire N11 complex in both H$\alpha$
and [O~{\sc iii}].  They also found multiple components, but concluded
that the main shell could best be described as undergoing a general
radial expansion at 45~km~s$^{-1}$, with lower density interior gas
undergoing more rapid, cellular expansion.

We now can find the luminosity predicted by the pressure-driven bubble
theory.  The typical emission measures of filaments around the edge of
the bubble are in the range 600--900 cm$^{-6}$~pc.
We take the expansion velocity to lie in the
range 45--60 km~s$^{-1}$.  To find the physical radius of the bubble,
we again use a distance to the LMC of 50 kpc (Panagia et al. 1991), so
that $1\arcmin = 14.5$~pc.  We then estimate the X-ray emitting bubble
to have a radius $R = $55--65 pc.  A pressure-driven bubble with these
parameter ranges is predicted by equation~(\ref{Lx}) to have an X-ray
luminosity $L_x = (0.6-1.2) \times 10^{35}$ erg s$^{-1}$.  This is a
factor of 3--7 lower than the X-ray luminosity derived from the
observations using constrained column densities: the main shell of N11
is somewhat X-ray bright.

The excess X-ray emission is most likely produced by an
intermittent source, since N11 represents an intermediate case between
the bubbles we described in Paper~III that show no excess,  and those we
described in Paper~I  with excesses of more than an order of
magnitude.  Arthur \& Henney (1996) suggested that an SNR going off
inside a bubble with embedded clumps could produce the excess.  They
ignored thermal evaporation of the clumps, suggesting that the
evaporation could be suppressed by magnetic fields.  \markcite{tao95}
Tao (1995) has shown, however, that fields cannot be tangled by
dynamical means sufficiently to strongly suppress evaporation, so
clumps would probably evaporate before being swept up by a SNR.
Alternatively, we have suggested in Paper~I that the excess X-ray
emission is being produced as interior SNRs expanding in the
low-density gas hit the ionized inner edge of the swept-up shell.

Using the same estimates of the observed parameters as above, we can
examine the physical parameters of this bubble using
equations~(\ref{n0}--\ref{mech}).  We find that the ambient density is
0.1--0.3~cm$^{-3}$, a reasonable interstellar value.  We note that
there are clearly regions of far higher density in the ring around the
main bubble, most noticeably traced by CO observations showing a ring
of molecular clouds roughly coincident with the regions of highest
H$\alpha$ intensity \markcite{cal96} \markcite{idg91}(Caldwell 1996;
Israel \& de Graauw 1991).  However, these regions appear to have a
low enough filling factor to not be dynamically important.

The mechanical luminosity driving the bubble (eq.~[\ref{mech}]) lies
in the range $L_{\rm mech} = (3-6) \times 10^{37}$~erg~s$^{-1}$.
Using only the 18 O-type stars in LH9 that have been spectroscopically
classified by Parker et al.\ (1992) and using the empirical mass loss 
rates and stellar wind terminal velocities appropriate for the spectral 
types (\markcite{con88}Conti \& Underhill 1988), we derive an integrated 
wind mechanical luminosity of 3.3$\times$10$^{37}$ erg s$^{-1}$.  It is
conceivable that the WC5-6 star in HD\,32228 and other unclassified
early type stars in LH9 could supply the $L_{\rm mech}$ required
to drive shell 1.

The total input energy $E = L_{\rm mech} t$ in such a bubble is then
(0.6--1.2)$ \times 10^{51}$~erg, just allowing the small number of supernovae
suggested by \markcite{ros96} Rosado et al.\ (1996) from detailed
modeling of the IMF.  Note that the possible hidden SNR producing the
excess X-ray emission would not yet have fully transferred its energy
to the shell, and so would currently make only a small contribution to
the quoted input energy.  The kinetic energy of the shell, from
equation~(\ref{ke}), is $K = (1-2) \times 10^{50}$ erg.  This model is
definitely inconsistent with the dozens of supernovae suggested by Meaburn et
al. (1989) \markcite{mea89} based on the overall SN rate in the LMC
and the size of the shell.

The age of the bubble is more problematic.  From equation~(\ref{t6})
and the above estimates of radius and expansion velocity, we find ages
of 0.5--0.8~Myr, well below the age of the central OB association.
Another way of looking at it is that this bubble is expanding much too
fast to have been produced by such an old OB association.
\markcite{oey96}Oey (1996) has demonstrated that this same phenomenon
occurs in a number of superbubbles where the age of the central OB
association has been measured.  A possible solution to this problem
may be that the expansion of the superbubble in a relatively low
density medium only occurs after it has broken out of its parent
molecular cloud, well after the formation of the central OB
association.  The association LH~10 in the nebula N11~B, which we
discuss next, may in fact be an example of just such an association
that has formed, but has not yet broken out into the low density
medium and produced a clear supershell.  This suggestion supposes that
the boundary between dense molecular cloud and diffuse ISM is
significantly steeper than the $r^{-2}$ power law that Oey (1996)
examined.  Further modeling of the process of superbubble breakout from
molecular clouds appears useful.

\subsection{N11B}

\label{n11b}
An unconstrained spectral fit to this region gives a
range in column density $\log N_{\rm H} =20.3-22.0$ and temperature
$kT=0.1-1.0$~keV.  Constraining
the absorbing column density independently, we find more reasonable
values of $\log N_{\rm H}=$21.3-21.6 and $kT= $0.4--0.5~keV.  This
corresponds to an X-ray luminosity $L_x =(0.3-0.8) \times
10^{35}$ erg s$^{-1}$.

In order to estimate the X-ray emission contributed by OB stars in the
central cluster LH~9, we first convert V band fluxes of the 24 O and B stars
located in a 4 arcmin$^2$ region centered on HD~32228 into bolometric
magnitude ($M_{\rm bol}$) by applying reddening \markcite{par92}
(Parker et al. 1992) and bolometric \markcite{cg91} (Chlebowski \&
Garmany 1991) corrections. We then convert the summed $L_{\rm bol}$ to
X-ray luminosity by the relation $L_x =(1.4 \pm 0.3)\times
10^{-7}L_{\rm bol}$, which is good for stars of types between O3 and
A5 \markcite{pal81} (Pallavincini et al.\ 1981; \markcite{chl89}
Chlebowski, Harnden, \& Sciortino 1989). The overall stellar X-ray
luminosity $L_x$ is 5$\times$10$^{33}$ erg s$^{-1}$, which is
6--17\% of the observed X-ray luminosity in N11~B.

Figures~\ref{echimage}(b) and \ref{echimage}(c) show two echellograms of the 
N11~B region.  The two slit positions are in a cross roughly centered on
N11~B, as shown in Figure~\ref{halpha}(c) and described in \S~\ref{ech}.
They show well-resolved, remarkably Gaussian line profiles, with FWHM
of 30--40~km~s$^{-1}$, far broader than the instrumental profile.
Figure~\ref{echn11b} shows an example of one of these profiles at one
of the few places where a high-velocity shoulder might be visible on
the line.  
Other than this shoulder, the lines appear quite featureless, and
completely lack high-velocity emission at velocities
greater than 100~km~s$^{-1}$.

The Carina Nebula resembles N11B in stellar content, both containing
several O3 stars.  A major difference between them is that the Carina
nebula shows an expanding shell structure in velocity and shows
high-velocity gas at velocity offsets of several hundred km s$^{-1}$
(\markcite{wal82}Walborn 1982; \markcite{wh82}Walborn \& Hesser 1982),
while N11B appears to have neither of these two dynamic features.  The
heating mechanism for the X-ray emitting gas in Carina could be either
stellar winds or SNRs.

The lack of a visible ring structure in N11B suggests that it contains
no large cavity of hot gas.  This makes it difficult to hide an SNR in
the region.  A supernova going off in a region with density greater
than 0.1 cm$^{-3}$ will generate classical SNR signatures
\markcite{tom81} (Tomisaka, Habe, \& Ikeuchi 1981).  We can find a
lower bound to the density in N11B by assuming that it is a uniform,
ionized sphere.  The emission measure in H$\alpha$ is $\sim 4\times
10^4$~cm$^{-6}$~pc, and the diameter of the region is 27 pc,
yielding an electron density of $n_e = 38$ cm$^{-3}$.  An SNR
expanding into gas this dense would produce fireworks, in the form of
the classical SNR signatures, none of which are observed. LH 10
contains between three and six O3 stars \markcite{par92} (Parker et
al. 1992), suggesting that the cluster is extremely young, and adding
further weight to the suggestion that no SNR has occurred yet.

If stellar winds do produce the observed X-ray emission, however, the
question arises of why no clear stellar wind bubbles can be seen in N11B.
A pressure-driven bubble would have a radius of \markcite{wea77}
(Weaver et al. 1977)
\begin{equation}
R = (28 pc) (\dot{M}_6 v_{2000}^2 n_0^{-1} t_6^3)^{1/5},
\end{equation}
where $\dot{M}_6$ is the mass-loss rate of the stellar wind in units
of $10^{-6} M_{\odot}/$yr, $v_{2000}$ is the stellar wind velocity in
units of 2000 km s$^{-1}$, and $t_6$ is the age of the bubble in units
of $10^6$ yr.  Substituting a density of $n_0 = 38$ cm$^{-3}$ and
setting the other values to unity gives a bubble of radius 14 pc, or
nearly an arcminute at the distance of the LMC.  Indeed,
\markcite{ros96} Rosado et al. (1996) suggest a structure of
overlapping bubbles based on their Fabry-Perot observations of the
region in H$\alpha$ and [O~{\sc iii}].  They found moderately broad
H$\alpha$ lines, consistent with our echelle observations, as might be
expected from many overlapping bubbles, and found that the narrower
[O~{\sc iii}] lines showed splitting indicative of small bubbles in
some regions.  Our H$\alpha$ image also shows some structure
suggestive of overlapping bubbles; this can be seen even more clearly
in the H$\alpha$ image of N11~B shown in \markcite{wp92} Walborn \&
Parker (1992).  We have been awarded time to perform high-resolution
imaging with the Hubble Space Telescope that may resolve this structure
more clearly.  AXAF imaging may reveal whether X-ray emission is
confined to the interiors of such small wind-blown bubbles.

\subsection{Other Shells}

Shell 2 lies on the north edge of N11~B.  In H$\alpha$ emission, the shell
appears elongated in the N-S direction, with a diameter of
approximately $5\arcmin$ in that direction.  However, the combination
of the X-ray image and echelle spectroscopy by \markcite{mea89}
Meaburn et al.\ (1989) shows a different picture.  X-rays are only
emitted in the southern third of the bubble (see Figure~\ref{hax}).
Spectral fits with constrained absorption give a $kT$ range 0.3--0.4~keV
and an X-ray luminosity of  $(0.95-1.1) \times 10^{35}$ erg
s$^{-1}$ in the 0.5--2.4~keV band.  This region also turns out to
show a line splitting of $\sim 20$ km~s$^{-1}$ in the echelle
spectrum.  The region without X-ray emission also shows no
line-splitting in H$\alpha$.  Thus the X-ray emission is associated
with a kinematically distinct structure of unknown origin.  

Shell 3 lies adjacent to Shell 2, north of the main bubble and NW of
N11~B.  A $kT$ range 0.1--0.5~keV and an X-ray luminosity range of
(0.7--0.9) $\times 10^{35}$ erg s$^{-1}$ are derived from the
constrained $N_{\rm H}$ fits.  There is no clear boundary between the
X-ray emission in the southern region of Shell~2 and that of Shell~3,
although the H$\alpha$ morphology suggests that they are separate
structures.  The X-ray emission from Shell~3 actually extends further
to the west, possibly even merging with the plume of emission north of the
SNR N11~L.  Spectroscopy of the region would be needed
to further understand the relation among these structures.

Shell 4 lies south of the main bubble.  Even the unconstrained fit
shown in Figure~\ref{gridplots} is quite close to the independently
derived constraints on \NH.  Spectral fits with constrained
absorption give a $kT$ range 0.3--0.5~keV and an
X-ray luminosity range of (0.97--1.2)$ \times 10^{35}$ erg s$^{-1}$.
It is perhaps most remarkable in that it appears quite similar in
H$\alpha$ morphology to the shell immediately to its E, and yet has
detectable X-ray emission, while the other shell does not.  In the
absence of any kinematical information from spectroscopy, we can only
suggest that intermittent events such as hidden supernovae might
produce such a pattern.

\subsection{HD~32228}

X-ray emission is observed in the PSPC map peaking at the position of
the bright cluster HD 32228 (RA 04$^{\rm h}56^{\rm m}34{\rlap.^{\rm
s}}$5; Dec $-66^\circ28'25''$; J2000) in the center of the main N11
shell (marked in Figure 1(a)).
However, no X-ray emission is seen in the HRI observation over 3
$\sigma$ or $2.3 \times 10^{-4}$ counts s$^{-1}$. This indicates that
the X-ray luminosity from any point source associated with HD 32228 is
below an upper limit of $L_{\rm x} \sim$ 4 $\times 10^{33}$ erg
s$^{-1}$.

The observed PSPC count rate within the 4 arcmin$^2$ region is $9.50
\times 10^{-4}$ counts s$^{-1}$.  If we take the absorbing column
density and local plasma temperature to be roughly the same as
elsewhere in Shell~1 ($kT \sim 3\times 10^{6}$ K, and $\log N_{\rm H}
\sim 20.5$, from Table~1), the observed X-ray luminosity is $L_x = 4.8
\times 10^{33}$ erg s$^{-1}$.  Following the same procedure used in
\S~\ref{n11b}, we estimate the stellar X-ray luminosity to be $L_*
\simeq (1.0 \pm 0.5) \times 10^{33}$ erg s$^{-1}$.  Therefore, the
stellar X-ray emission accounts for only $\sim$ 30\% of the observed
$L_x$, and the remaining emission is from the hot gas.  HD~32228 does
not appear to contain a compact object similar to those observed near
R136 and R140 in 30 Dor (Wang 1995), showing that such X-ray sources
are not universally associated with tight clusters such as this.

\subsection{NW region}

The H$\alpha$ image shows faint filaments extending above Shell~2 to
the NW, associated with diffuse X-ray emission, as well as with an
unidentified X-ray point source, all lying within the box labeled
NW~diffuse in Figure~\ref{hax}.  The spectrum of the point source
suggests it is a Galactic foreground star.  Its point nature has also
been verified by its spatial profile in the HRI image.  Therefore,
X-ray emission from the point source is subtracted from the total PSPC
flux.  Spectral fits to the remaining emission with constrained
absorption give a $kT$ of 0.3~keV in this region, and an X-ray
luminosity range of $(0.5-0.6) \times 10^{35}$ erg s$^{-1}$.
Similar diffuse X-ray emission has been observed outside the
major structures visible in H$\alpha$ in N44
(\markcite{paperII}Paper~II) and 30~Dor (\markcite{chu96}Chu 1996).

\subsection{N11 C}

The X-ray emission in this region is dominated by a weak, point-like
source.  Spectral fitting suggests a temperature of around 1 keV and a
low absorption column density of $\log N_{\rm H} = 20.4$.  These
properties are consistent with a galactic stellar source in the solar
neighborhood.  Unfortunately, we cannot confirm the point nature of
this source, as it has luminosity below the detection limit of our HRI
observation.

\section{Conclusions} %5

1. The main bubble of N11, surrounding the OB association LH 9, has an
X-ray luminosity of $(3-4) \times 10^{35}$ erg s$^{-1}$, a factor of
3--7 times higher than the prediction of an energy-conserving,
pressure-driven superbubble model.  This could be the result of an SNR
hitting the inner edge of the shell (Paper~I), or of a mass-loaded
SNR in the interior \markcite{ah96} (Arthur \& Henney 1996).  Both of
these processes should produce intermittent and time-dependent
excesses of X-ray emission over the prediction of the pressure-driven
model.  Indeed, the X-ray emission from the main bubble of N11 is only
slightly above the predicted value, and the bubbles of Paper III all have
upper limits on their luminosities consistent with the prediction.
It remains true that no superbubble has yet been positively detected
emitting with luminosity within a factor of three of the prediction of
the pressure-driven model, although the limits are narrowing.

2. Diffuse X-ray emission is detected from N11~B, a region analogous
to the Carina nebula with multiple O3 and other massive stars.  The
luminosity greatly exceeds that expected from the stars themselves, as
in Carina.  Although no single bubble is observed, echelle
observations show strongly supersonic motions, H$\alpha$ images show
hints of overlapping shell structures, and the Fabry-Perot
observations of [O~{\sc iii}] by Rosado et al.\ (1996) show line
splitting in some regions.  No gas moving at velocities over
60~km~s$^{-1}$ is seen however, suggesting a lack of SNRs.  Thus we
conclude that the diffuse X-ray emission could be powered solely by 
stellar winds from the central massive stars of LH~10. 

The application to Carina is, unfortunately, less than incisive.
An absence of emission from N11B in excess of stellar emission would
have been strong support for the suggestion that hidden SNRs produce
the X-ray emission in Carina, because the clusters are otherwise
similar.  However, the presence of bright X-ray emission from N11B
does not argue against hidden SNRs in Carina, because of the presence
of high-velocity gas in Carina and the absence of such gas in N11B.

3. The dynamical age of the main shell of N11 is under 1 Myr,
significantly shorter than the age derived by Parker et al. (1992)
\markcite{par92} for the central cluster, LH~9.  This same problem has been
found for other superbubbles by Oey (1996).  We propose that the
solution to this problem is that during the initial evolution of OB
associations, they remain cocooned within their high-density parent
molecular clouds, and do not begin generating the observed
superbubbles until after they have broken out of the dense clouds into
the surrounding diffuse ISM.  N11~B may be an example of a young OB
association with energetic winds that has not yet broken out of a
high-density region, since despite the presence of O3 stars and X-ray
emission, very little high-velocity gas is observed, while CO is still
observed towards this region.

4.  The tight cluster HD~32228 in LH~9 is not a strong X-ray source,
despite its resemblance in stellar density and size to the clusters
R136 and R140 in 30 Dor.  This suggests that the strong X-ray emission
observed from R136 and R140 is not an intrinsic property of tight
stellar clusters, but is rather generated by a compact object that has
formed there, as suggested by \markcite{wan96}Wang (1995).

\acknowledgments We thank G. Garc\'{\i}a-Segura for participation in
early stages of this project.  SDP acknowledges the hospitality of the
Max-Planck-Institut f\"ur Astronomie, and support from a National
Science Foundation Graduate Fellowship.  This research was partially
supported by NASA grants NAG5-2245 and NAG5-1900.

\clearpage

%\begin{figure}[hp]  %1
\begin{center} {\large \bf Figure Captions} \end{center}

%\label{halpha}
\figcaption{(a) Green continuum image of N11 from the Curtis
Schmidt, showing the
distribution of stars.  The OB associations and HD~32228 are marked.
(b) H$\alpha$ Schmidt image of N11 to show the distribution of ionized
gas.  The OB association LH~9 lies in the center of the main shell;
LH~10, LH~13, and LH~14 lie in N11~B, N11~C, and N11~E, respectively.
The compact cluster HD~32228 lies just east of the center of the 
main shell, at RA 04$^{\rm h}$56$^{\rm m}$34\rlap.$^{\rm s}$5, 
Dec -66$^{\circ}28'25''$ (J2000).  The three echelle slit positions
are marked.
\label{halpha}}
%\end{figure}

%\begin{figure}[hp]  %2
\figcaption{X-ray image of the N11 region.  The PSPC image was binned
by a factor of 10 and smoothed with a gaussian of $\sigma$ = 25$''$.
The RA and Dec are in J2000.  The contours are at 10\%, 15\%, 20\%, 25\%,
and 30\% of the peak brightness of the supernova remnant N11L at 
04$^{\rm h}$54$^{\rm m}$48\rlap.$^{\rm s}$5, -66$^{\circ}25'45''$.
\label{rawx}}
%\end{figure}

%\begin{figure}[hp]  %3
\figcaption{Contours of X-ray emission superimposed on our Curtis
Schmidt H$\alpha$ image.  This X-ray image was derived by binning
the PSPC image by a factor of 8 and smoothed by a gaussian of $\sigma$ 
= 8$''$.  The contours are 10\%, 13\%, 16\%, and 20\% of the
peak brightness of the supernova remnant N11L at 
04$^{\rm h}$54$^{\rm m}$48\rlap.$^{\rm s}$5, -66$^{\circ}25'45''$.
The RA and DEC are in J2000.
\label{hax}}
%\end{figure}

%\begin{figure}[hp]  %4
\figcaption{Echellograms of N11. (a) E-W oriented slit centered on
Sk -66$^\circ$31 close to the center of the main shell.  (b) E-W
oriented slit 244\arcsec~N, in N11~B, centered on the star PGMW~3021. 
(c) N-S oriented slit also crossing the star PGMW~3021 in N11~B. The
profiles shown in Figures~\ref{echmain} and~\ref{echn11b} are drawn
from the regions bounded by the thin black lines.  Unvignetted slit
lengths are 3\farcm5. \label{echimage}}
%\end{figure}

\figcaption{Observed spectral energy distributions and the best model
fits {\em without} constraints on \NH.
\label{x-spec}} %5

%\begin{figure}[hp]  %6
\figcaption{$\chi^2$ grid plots for the spectral fits shown in
Figure~\ref{x-spec}.  The three
contours correspond to 68\%, 90\%, and 99\% confidence levels.
Temperatures are given in keV.
\label{gridplots}}
%\end{figure}

%\begin{figure}[hp]  %7
\figcaption{Velocity profiles at the four places designated by the
thin black lines in Figure~\ref{echimage}(a), moving from left to right.
\label{echmain}}
%\end{figure}

%\begin{figure}[hp]  %8
\figcaption{Velocity profile at the place designated by the
thin black lines in Figure~\ref{echimage}(c). 
\label{echn11b}}
%\end{figure}

\clearpage
\begin{deluxetable}{lllllll}
\tablewidth{0pt}    %natural size rather than text size
\tablecaption{Observed X-ray properties of N11\tablenotemark{b}
      \label{spectra}}
%\tablehead{\colhead{\bf regions}  \colhead{Shell 1} \colhead{Shell 2}
%\colhead{NW diffuse} \colhead{Shell 3} \colhead{Shell 4}
%\colhead{N11~B} } 
\tablehead{{\bf Regions} & Shell 1 & Shell 2 & NW diffuse & Shell 3 & Shell
	4 & N11 B  } 
\startdata 

\sidehead{\bf Free N$_{H}$ fits}

$kT$ (keV)     & 0.1--0.4 & 0.1--0.5 & 0.1--0.6 &  0.1--0.5 
	     & 0.3--0.5 & 0.1--1.0  \nl

$\log N_{\rm H}$ & 20.5--22.0 & 20.4--22.0 & 20.0--22.0 & 20.2--22.0 
	     & 20.0--20.8 & 20.3--22.0 \nl

$\log L_{x}$ (0.1--2.4)\tablenotemark{a} & 35.57--39.36 &
35.00--38.85& 34.61--38.52 & 34.78--38.65 & 34.92--35.28 & 34.39--38.26 \nl
$\log L_{x}$ (0.5--2.4)\tablenotemark{a} & 35.39--38.11 & 34.85--37.60 
& 34.46--37.27 & 34.62--37.40 & 34.77--34.98 & 34.29--37.01 \nl 

\sidehead{\bf Fixed $N_{\rm H}$ fits}

$kT$ (keV)     & 0.31--0.33 & 0.34--0.36 & 0.30--0.32 & 0.27--0.29 
	     & 0.26--0.27 & 0.36--0.51  \nl

$\log N_{\rm H}$ & 20.9--21.1 & 20.9--21.1 & 20.9--21.1 & 20.9--21.1 
	     & 20.9--21.1 & 21.3--21.6 \nl

$\log L_{x}$ (0.1--2.4)\tablenotemark{a} & 35.73--35.84 & 35.18--35.28 &
34.90--35.02 & 35.09--35.22 & 35.27--35.38 & 34.69--35.09 \nl
$\log L_{x}$ (0.5--2.4)\tablenotemark{a} & 35.51--35.60 & 34.98--35.07 &
34.67--34.77 & 34.83--34.94 & 34.99--35.09 & 34.53--34.89 \nl
\enddata

\tablenotetext{a}{In units of  erg~s$^{-1}$, with the energy band
given in keV.}
\tablenotetext{b}{Parameter ranges give 99\% confidence levels.}

\end{deluxetable}

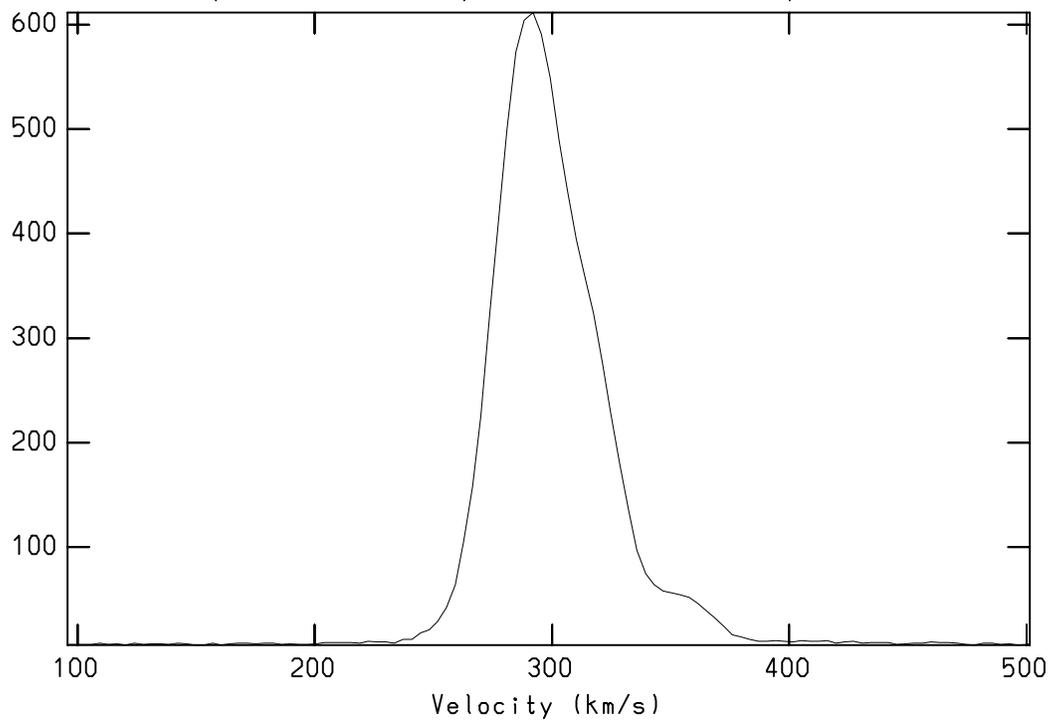
\end{document}